\documentclass[iop]{emulateapj}

\def\lapp{\ifmmode\stackrel{<}{_{\sim}}\else$\stackrel{<}{_{\sim}}$\fi}
\def\gapp{\ifmmode\stackrel{>}{_{\sim}}\else$\stackrel{>}{_{\sim}}$\fi}

\usepackage{natbib}
\usepackage{graphicx}
\usepackage{float}

\newcommand{\psr}{PSR~J0337+1715}

\submitted{Submitted to ApJL October 31, 2013; Accepted November 29, 2013}

\shorttitle{Formation of a Triple Millisecond Pulsar}
\shortauthors{Tauris \& van~den~Heuvel}

\begin{document}

\title{Formation of the Galactic Millisecond Pulsar Triple System PSR~J0337+1715 \\
       -- a Neutron Star with Two Orbiting White Dwarfs}

\author{
T.~M.~Tauris\altaffilmark{1,2} and E.~P.~J. van~den~Heuvel\altaffilmark{3}
}

\altaffiltext{1}{Argelander-Institut f\"ur Astronomie, Universit\"at Bonn, Auf dem H\"ugel 71, D-53121 Bonn, Germany}
\altaffiltext{2}{Max-Planck-Institut f\"ur Radioastronomie, Auf dem H\"ugel 69, D-53121 Bonn, Germany}
\altaffiltext{3}{Astronomical Institute Anton Pannekoek, University of Amsterdam, P.O. Box 94249, 1090~GE Amsterdam, The~Netherlands}

\begin{abstract}
The millisecond pulsar in a triple system (\psr, recently discovered by Ransom~et~al.) 
is an unusual neutron star with two orbiting white dwarfs. The existence of such a system in the Galactic field poses new challenges to stellar astrophysics for
understanding evolution, interactions and mass-transfer in close multiple stellar systems.
In addition, this system provides the first precise confirmation for a very wide-orbit system of the white dwarf mass$-$orbital period relation.
Here we present a self-consistent, semi-analytical solution to the formation of \psr. Our model constrains the peculiar velocity of the system to be
less than $160\;{\rm km}\,{\rm s}^{-1}$ and brings novel insight to, for example, common envelope evolution in a triple system, for which we find
evidence for in-spiral of both outer stars. Finally, we briefly discuss our scenario in relation to alternative models.
\end{abstract}

\email{tauris@astro.uni-bonn.de}

\section{Introduction}
\label{sec:intro}
Stars are possibly always formed in multiple systems \citep[e.g.][]{bbv03} and 
observational estimates suggest that about 20\%--30\% of all binary stars are in fact members of triple systems \citep{ttsu06,rdl+13}. 
Triple systems can remain bound with a long-term stability if they have a hierarchical structure (e.g. a close inner binary with a third star in relatively distant orbit).
In addition, a number of peculiar binary pulsars have recently
been discovered, such as PSR~J1903+0327 \citep{crl+08}, which require a triple system origin \citep[e.g.][]{fbw+11,pvvn11,pcp12}.

The discoveries of binaries with a triple origin is not unexpected. \citet{it99} estimated that in 
$\sim$70\% of the triple systems, the inner binary is close enough that the most massive star will evolve to fill its Roche~lobe. 
Furthermore, in $\sim$15\% of the triples, the outer third (tertiary) star may also fill its
Roche~lobe at some point, possibly leading to disintegration or production of rare configurations with three degenerate objects in the same system.
Recently, \citet{rsa+14} have reported the discovery of \psr, which is the first example of such an exotic system
-- a neutron star orbited by two white dwarfs. 

\psr\ is a triple system located at a distance of $\sim\!1.3\;{\rm kpc}$. It contains a $1.438\;M_{\odot}$ radio millisecond pulsar (MSP)
with a spin period of $P=2.73\;{\rm ms}$ and two white~dwarfs (WDs) with 
masses of $M_{\rm WD,2}=0.197\;M_{\odot}$ and $M_{\rm WD,3}=0.410\;M_{\odot}$, and orbital periods of 
$P_{\rm orb,12}=1.63\;{\rm days}$ and $P_{\rm orb,3}=327\;{\rm days}$, respectively. 
Thus this triple system is highly hierarchical with a close inner binary and a distant tertiary star.
In addition, the system is almost exactly coplanar ($\delta _i=0.01^\circ$), and the orbits are quite circular with 
eccentricities of $e_{12}=6.9\times 10^{-4}$ and $e_3=0.035$ \citep{rsa+14}. 

Here, we investigate the formation of such a triple compact object system and present a model which aims to explain and reconcile the observed data with 
current theories of stellar interactions.

\section{Progenitor Evolution of \psr}\label{sec:triple_evol}
\begin{figure}[t!]
    \centering
    \includegraphics[width=1.00\columnwidth,angle=0]{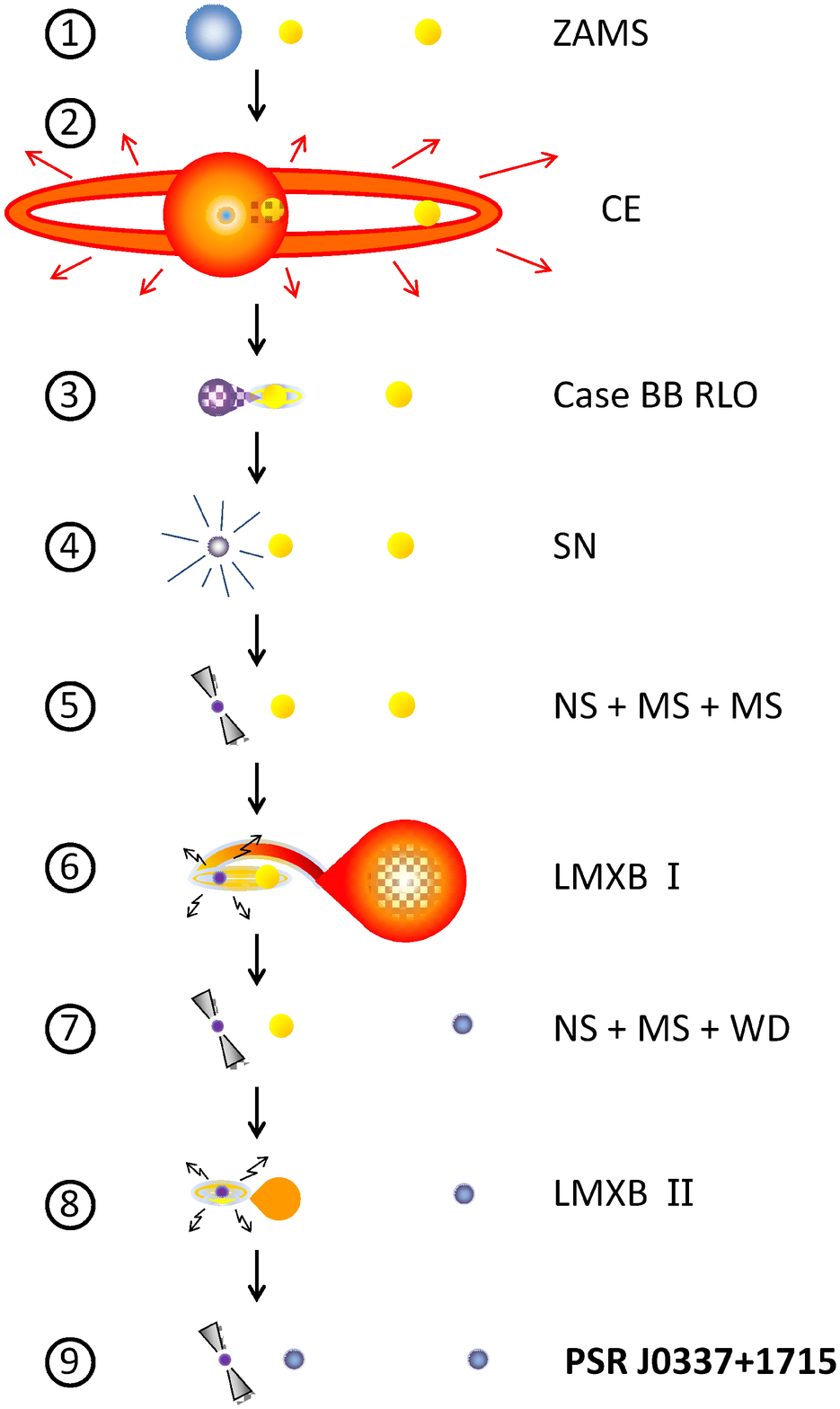}
    \caption{Illustration of our triple star evolution from the zero-age main
        sequence (ZAMS) to the present observed system \psr. Numerical parameters are given in Table~\ref{table:stages}.
        The initially massive B-star evolves to initiate Roche-lobe overflow (RLO) towards the inner G/F-star, leading to
        dynamical unstable mass transfer and the formation of a common envelope (CE), partially embedding the outer F-star. 
        The resulting helium star (the naked core of the massive star) expands and initiates another phase of (Case~BB) RLO, before it 
        collapses into a neutron star (NS) in a supernova (SN) explosion. Thereafter, the system becomes visible as a young radio pulsar
        with two main-sequence (MS) stars. Given that the tertiary star is more massive than the secondary star, the outer LMXB~phase
        (lasting $15-20\;{\rm Myr}$) occurs before the inner LMXB~phase. The latter mass-transfer episode proceeds on a long timescale 
        ($2\;{\rm Gyr}$), causing the NS to become a fully recycled MSP when finally orbited by two white dwarfs (WDs).
        \label{fig:cartoon}}
\end{figure}

To investigate the formation of \psr\ we start with constraints obtained from the present-day triple system and trace
the evolution backward.
Before elaborating on the details, we briefly summarize the outline of our model which
is illustrated in Figure~\ref{fig:cartoon}. Numerical parameters are provided in Table~\ref{table:stages}. 

\begin{center}
\begin{deluxetable*}{lllrrrrrrrrr}
\tablecaption{Triple System Parameters at the Onset of each Stage in Our Scenario (Figure~\ref{fig:cartoon}) for the Formation of \psr\ \label{table:stages}}
\tablewidth{0pt}
\startdata        
        \hline
        \hline
	\noalign{\smallskip}
	\noalign{\smallskip}
         & & Stage: & 1 & 2 & 3 & 4 & 5 & 6 & 7 & 8 & 9\\
	\noalign{\smallskip}
        \hline
	\noalign{\smallskip}
         Age of the system               & $t$              & $[{\rm Myr}]$                & 0.0  & 23.3 & 23.3 & 25.1 & 25.2 & $5\,500$ & $5\,517$ & $8\,500$ & $10\,500$\\
         Mass of primary star            & $M_1$            & $[M_{\odot}]$                & 10.0 & 9.90 & 2.90 & 1.70 & 1.28 & 1.28     & 1.30     & 1.30     & 1.438\\
         Mass of secondary star          & $M_2$            & $[M_{\odot}]$                & 1.10 & 1.10 & 1.10 & 1.10 & 1.10 & 1.10     & 1.12     & 1.12     & 0.197\\
         Mass of tertiary star           & $M_3$            & $[M_{\odot}]$                & 1.30 & 1.30 & 1.30 & 1.30 & 1.30 & 1.30     & 0.410    & 0.410    & 0.410\\
         Orbital period of inner binary  & $P_{\rm orb,12}$ & $[{\rm days}]$               & 835  & 849  & 2.47 & 0.95 & 1.55 & 1.55     & 1.50     & 0.90     & 1.63\\
         Orbital period of tertiary star & $P_{\rm orb,3}$ & $[{\rm days}]$                & 4020 & 4080 & 17.1 & 15.7 & 15.3 & 14.2     & 250      & 250      & 327\\
         Eccentricity of inner binary    & $e_{12}$ &                                      & 0.00 & 0.00 & 0.02 & 0.01 & 0.24 & 0.20     & 0.02     & 0.00     & 0.00\\
         Eccentricity of outer orbit     & $e_{3}$  &                                      & 0.00 & 0.00 & 0.04 & 0.04 & 0.22 & 0.03     & 0.03     & 0.03     & 0.03\\
         Stability parameter             & $(R_{\rm peri}/a_{\rm in})$ &                   & 2.96 & 2.96 & 3.83 & 7.07 & 4.15 & 4.89     & 30.8     & 43.3     & 35.6\\
         Critical stability limit        & $(R_{\rm peri}/a_{\rm in})_{\rm crit}$ &        & 2.93 & 2.93 & 3.21 & 3.44 & 3.93 & 3.80     & 3.04     & 3.04     & 3.13\\
         Temperature of outer WD         & $T_{\rm eff,3}$  & $[{\rm K}]$                  &      &      &      &      &      &          & 18\,000  & 5\,800   & 4\,300\\
	\noalign{\smallskip}
        \hline
        \hline
\enddata
\end{deluxetable*}
\end{center}

\subsection{Summary of Our Model}\label{subsec:outline}
According to our model, 
the system started out on the zero-age main sequence (ZAMS) with a roughly $10\;M_{\odot}$ primary star and two companions with masses of about $1.10\;M_{\odot}$ and $1.30\;M_{\odot}$, 
for the secondary and the tertiary star, respectively (Table~\ref{table:stages}, stage~1).
After a common envelope (CE)~phase (stage~2) where the extended envelope of the primary engulfed the other two stars 
(initially only embedding the secondary star; later also partly the tertiary star), 
the orbital period of the inner system was $P_{\rm orb,12}=2.47\;{\rm days}$ and the orbital period of the outer star was $P_{\rm orb,3}=17.1\;{\rm days}$. 
Following a second mass transfer (Case~BB, stage~3) and a supernova (SN)~explosion (stage~4) they became 
$P_{\rm orb,12}=1.55\;{\rm days}$ and $P_{\rm orb,3}=15.3\;{\rm days}$, 
which after orbital circularization (before stage~6) became $P_{\rm orb,12}=1.55\;{\rm days}$ and $P_{\rm orb,3}=14.2\;{\rm days}$. 
The last set of values were the orbital periods at the onset of the first (outer) 
low-mass X-ray binary (LMXB)~phase, which ended with $P_{\rm orb,12}=1.50\;{\rm days}$ and $P_{\rm orb,3}=250\;{\rm days}$,
before the second (inner) LMXB~phase left the system with its present observed properties. 
We now describe in more detail the physical properties of our model.

\subsection{The $M_{\rm WD}-P_{\rm orb}$ Relation}\label{subsec:MwdPorb-rel}
\begin{figure*}[t]
\centering
\includegraphics[width=1.20\columnwidth,angle=-90]{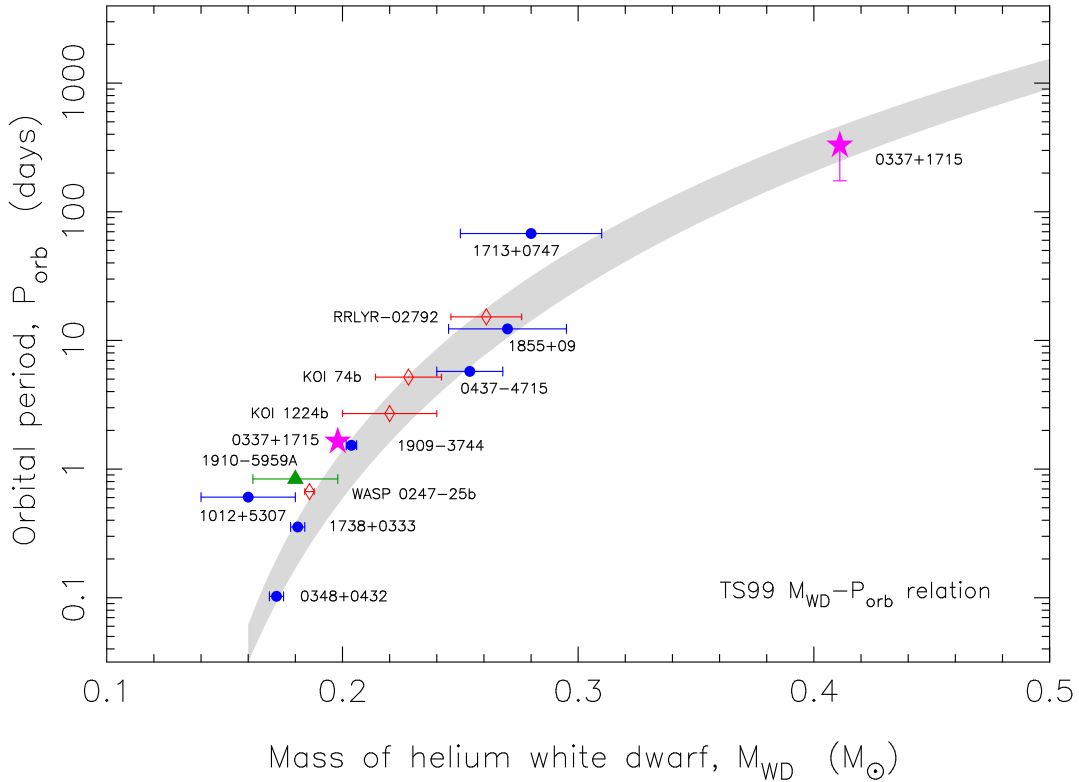}
\caption{
  $M_{\rm WD}-P_{\rm orb}$ relation (TS99), as calculated by \citet{ts99}. The width of the relation is caused by using metallicities from
  $Z=0.001-0.02$. Observational data is plotted for helium WD companions orbiting pulsars, for which the $1\sigma$~uncertainties are less
  than 10\% of $M_{\rm WD}$. Also included are four Galactic field proto-WDs orbiting an A-type main sequence star 
  (WASP~0247-25b, KOI~1214b, KOI~74b and RRLYR$-$02792). The discovery of the triple MSP J0337+1715 
  adds two important high-precision data points to this graph and strengthens the validity of the relation. 
  The error~bar plotted for the outer WD is caused by an uncertainty in the widening of the outer orbit during the inner LMXB~phase, see text. 
  Theoretically, the relation becomes uncertain for $P_{\rm orb}<1\;{\rm day}$.
  \citep[For references to data, with increasing $P_{\rm orb}$, see:][]{afw+13,avk+12,vbjj05,msm+13,cbp+12,jhb+05,rsa+14,brvc12,vrb+10,vbv+08,spa04,ptg+12,sns+05,rsa+14}.
  }
\label{fig:MPrel}
\end{figure*}

A close triple system like \psr\ with one neutron star (NS) and two WDs requires two LMXB~phases. 
Although \psr\ was substantially less hierarchical earlier in its evolution, the system 
seems to have evolved through both of its LMXB~phases, in~effect, mainly via binary interactions, 
with only small dynamical perturbations from the second or third star.
The important piece of evidence for this comes from the masses and orbital periods of the WDs which fall exactly
as predicted by the $M_{\rm WD}-P_{\rm orb}$ relation for LMXB evolution \citep[e.g.][]{sav87,rpj+95,ts99,vbjj05}. 
The match between this theoretical relation and the observational data for \psr\ is excellent.
This is demonstrated in Figure~\ref{fig:MPrel} where we plot all available data of helium~WDs with 
masses measured to an accuracy $1\sigma < 0.1\;M_{\rm WD}$. These helium~WDs are companions to pulsars
or found in binaries with A-type main sequence stars. 

The eccentricities of both orbits are, although small, one to two orders of magnitude larger than expected 
theoretically for isolated binaries with similar components and orbital periods \citep{pk94}. Although this may be a result of
mutual triple interactions, a few binary pulsars with WD companions in the Galactic field have similar eccentricities \citep[see Figure~4 in][]{tlk12}.

\subsection{Evolution of the Two LMXB~Phases}
The two WDs orbiting \psr\ are the remnants of two LMXB~phases.
Optical observations by \citet{kap+14} show that
the inner WD is quite hot ($15\,800\pm 100\;{\rm K}$), whereas the outer WD is too cold to be detected. Therefore, we assume in the following
that the inner WD formed last (see Section~\ref{subsec:alternative} for a discussion). 

We can deduce that both LMXB~phases evolved highly non-conservatively since the low MSP mass of $1.438\;M_{\odot}$ implies that
it cannot have accreted much material (at most $0.1-0.2\;M_{\odot}$ in total; in our model we assume a NS birth mass of $1.28\;M_{\odot}$). 
Since the presently observed (post-LMXB) orbital period of the inner binary, $P_{\rm orb,12}=1.63\;{\rm days}$ 
is close to the so-called bifurcation period \citep[between 1 and~2~days,][]{ps89,ml09}, below which magnetic braking is dominant in LMXBs \citep{rvj83}
and above which the widening of the orbit is significant, we also conclude that the pre-LMXB (post-SN) orbital period of the inner binary 
must have been close to this value. 

For the preceding outer LMXB~phase there is further evidence for highly non-conservative evolution since the mass-transfer rate must have been super-Eddington 
for such a wide system (initially $P_{\rm orb,3}=14.2\;{\rm days}$) where the donor star had a deep convective envelope at the onset of the Roche-lobe overflow \citep[RLO;][]{ts99,prp02}. 
During the rapid outer LMXB~phase, we expect the pulsar only to be mildly recycled -- possibly with a spin period between 
25~ms and 1~sec. \citep[a typical value for pulsars with a WD and $P_{\rm orb}>200\;{\rm days}$,][]{tlk12}.
Effective recycling of the MSP was obtained in the subsequent long-lasting inner LMXB~phase.

The masses of the two WD progenitor stars are constrained, on the one hand, by requirements of the development of a degenerate
helium core {\it and} dynamically stable mass transfer ($M_{2,3}^{\rm ZAMS}\le 1.6\;M_{\odot}$) and, on the other hand, by nuclear evolution and WD cooling
within a Hubble time ($M_{2,3}^{\rm ZAMS}> 1.0\;M_{\odot}$).
The estimated component masses before/after the mass transfer then yield the amount of mass lost from the system. 

The changes of the orbital separations as a result of LMXB mass transfer/loss can be found by solving the orbital angular momentum balance equation 
within the isotropic re-emission model \citep{spv97}, but with 
some modification. For example, we cannot assume a pure fast (Jeans mode) wind mass loss from the inner binary with respect to the tertiary star during the inner LMXB~phase (stage~8). 
Some of the inner binary material might be lost in a rather slow wind which only causes a moderate 
widening (if any) of $P_{\rm orb,3}$ during this phase. 
Nevertheless, following the outer LMXB~phase (stage~6), $P_{\rm orb,3}$ could have been smaller than observed today (327~days). In the most extreme case
(Jeans mode) it could be as short as 175~days, given that the semi-major axis of the outer orbit in this case changes 
according to $a_{3f}=a_{3i}\,(M_i/M_f)$, where $M$ is the total mass of the triple system and the indices $i$ and $f$ refer to
initial and final values, respectively.
Here we assume a more moderate value of 250~days.

The major uncertainties in our modelling are related to i) spin-orbit couplings mediated by tidal torques (e.g. magnetic braking) 
and ii) accretion onto the inner binary system during mass transfer from the tertiary star (stage~6), and the poorly known
specific orbital angular momentum of the ejected mass.
Presumably, an inner circumbinary disk \citep{dijv13} will be formed which may subsequently influence the evolution of the inner orbit. 
However, investigating the dynamical effects of the SN explosion helps to constrain the properties of the pre-LMXB systems.

\subsection{The Dynamical Effects of the SN Explosion}\label{subsec:SN}
\begin{figure}
\centering
\includegraphics[width=0.70\columnwidth,angle=-90]{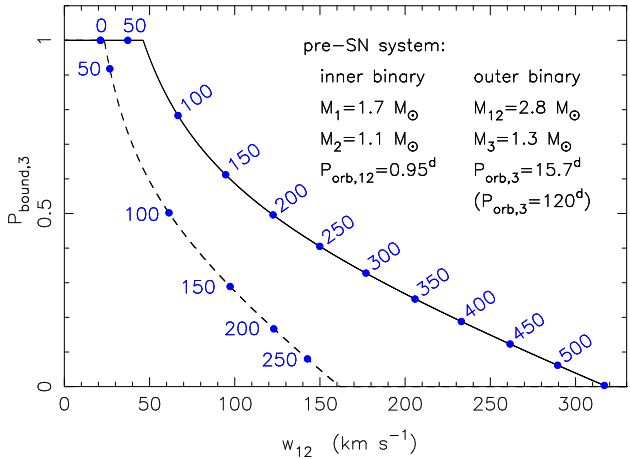}
\caption{
  Probability for a triple system to survive a given recoil velocity, $w_{\rm 12}$ obtained by the inner binary due to a SN.
  The bullet points represent average values of $w_{\rm 12}$ for the stated kick magnitudes  
  (in ${\rm km}\,{\rm s}^{-1}$) imparted on a newborn $1.28\;M_{\odot}$ NS in the inner binary.
  The kick directions were chosen from an isotropic distribution and simulations were done for random orbital phases between the mutual orbits.
  The solid (dashed) line is for a pre-SN $P_{\rm orb,3}=15.7\;{\rm days}$ ($P_{\rm orb,3}=120\;{\rm days}$).
  When calculating $\langle w_{\rm 12} \rangle$ only triple
  systems surviving the SN in long-term dynamically stable orbits, and for which the inner binary avoided merging, were considered.
  The assumed pre-SN core mass is $M_1=1.7\;M_{\odot}$. All other relevant pre-SN triple parameters are given in the figure. 
  }
\label{fig:kick}
\end{figure}
A general discussion of dynamical effects of asymmetric SNe in hierarchical multiple star systems is found in \citet{pcp12}, and references therein.
Here we have simulated the dynamical effects of the SN explosion that created the NS in the \psr\ system. 
In Figure~\ref{fig:kick} we have plotted the survival probability of our best model for the triple system as a function of the recoil velocity immediately imparted to
the inner system (i.e. the "inner binary kick", $w_{\rm 12}$) as a consequence of the SN. All relevant pre-SN parameters are stated in 
the figure (see also Table~\ref{table:stages}, stage~4). 
Along the plotted curves are examples of average values of $w_{12}$ for
kick velocities between $0$ and $550\;{\rm km}\,{\rm s}^{-1}$ which were imparted to the newborn NS. For estimating the resulting values of $w_{12}$
we only considered systems which survived the SN in long-term stable orbits (Section~\ref{subsec:stability}), and for which the 
inner binary avoided merging\footnote{The plotted probabilities do not take into account the specific requirements on the value of the post-SN $P_{\rm orb,3}$
necessary for forming \psr. If including this specific constraint, the probabilities shown would be much lower.}.

Given the constraints on the post-SN evolution to meet the requirements for \psr, we find that it may even have survived a NS kick up to
$400\;{\rm km}\,{\rm s}^{-1}$. The resulting peculiar velocities of the triple system range between $15$ and $160\;{\rm km}\,{\rm s}^{-1}$.

\subsection{Pre-SN Core Mass and Case~BB RLO}\label{subsec:pre-SN}
The triple system is much more likely to survive the explosion if the pre-collapsing core mass is low. A pre-SN core mass of only $\sim\!1.7M_{\odot}$ 
is indeed expected if the progenitor star (say, $M_1=10\;M_{\odot}$) lost its hydrogen-rich envelope on the red-giant branch (RGB), 
because the resulting naked helium core itself ($\sim\!2.9\;M_{\odot}$) would expand and give rise to Case~BB RLO \citep{hab86a},
leaving an even further stripped pre-SN core during stage~3. 
In case the NS progenitor did not lose its envelope until the asymptotic giant branch, the collapsing core mass could have been much larger. However, we find that a
collapsing core mass of, for example, $3.2\;M_{\odot}$ decreases the survival probability considerably and leaves the triple system
with $v\sim 150\;{\rm km}\,{\rm s}^{-1}$. 

\subsection{The Common Envelope Evolution -- New Lessons from a Triple System}\label{subsec:CE}
The outcome of the CE evolution is crucial for determining the pre-SN core mass and orbital periods \citep[e.g.][]{td01}, and thus for the survival probability of the triple system. 
Unfortunately, CE evolution is the least understood of the important interactions in close binary systems \citep[see][for a recent review]{ijc+13}. 
For close triple systems, understanding the CE evolution is an even more complicated task. However, the existence of \psr\ provides an important piece of information: 
namely that CE evolution (stage~2) not only leads to efficient orbital angular momentum loss of the inner binary orbit, also the tertiary star is subject to efficient in-spiral.  
The evidence for this conclusion is the following.
On the one hand, the orbital period of the tertiary star could not be very large at the moment of the SN. There are two reasons for this: 
i) the post-SN $P_{\rm orb,3}$ (after recircularization) must match the expected orbital period at the onset of the outer LMXB~phase, and 
ii) to avoid a very small survival probability as a consequence of the SN.
On the other hand, the system must have had $P_{\rm orb,3}\ga 4000\;{\rm days}$ on the ZAMS. The evidence for this is that the ratio of the
orbital periods ($P_{\rm orb,3}/P_{\rm orb,12}$) on the ZAMS must have been at least a factor of $\sim\!5$, and even larger for non-circular orbits, in order for the triple system
to remain dynamically stable on a long timescale (see Section~\ref{subsec:stability}). 
Furthermore, the onset of the CE could not have happened much earlier than near the tip of the RGB of the primary star, which corresponds to $P_{\rm orb,12}\ga 800\;{\rm days}$. 
The reason is that the binding energy of the hydrogen-rich envelope is simply too high to allow for ejection at earlier stages \citep{dt00}. 
This constraint, in combination with the stability criteria of the triple system, sets the lower limit of $P_{\rm orb,3}\sim 4000\;{\rm days}$ on the ZAMS. 
Hence, we conclude that an efficient in-spiral of the tertiary star, and thus {\it both} of the outer stars, must have taken place.  

\citet{pvvn11} argued for a similar conclusion based on the tertiary F-dwarf orbiting the LMXB 4U~2129+47 (V1727~Cyg). As also pointed out by these authors, the SN explosion itself 
could also have decreased the orbital period of the tertiary star of the surviving triple system. In that case, the need for CE in-spiral is less extreme, but still
highly demanded. From our simulations, we find that the pre-SN orbital period of the tertiary could have been up to about 120~days; still much shorter
than the ZAMS $P_{\rm orb,3}$ of about 4000~days.

\section{Discussion}\label{sec:discussion}
\subsection{Long-term Stability of a Triple System}\label{subsec:stability}
Throughout our scenario, we have checked at each evolutionary stage that the mutual orbits
of our solutions are expected to have a long-term dynamical stability. A number of
stability criteria for triple systems have been proposed over the last four decades \citep[see][for an overview]{mik08}.
To be extra cautious and conservative, we only accepted our found solutions in case they fulfilled all criteria
suggested by \citet{har72,bai87,ek95,ma01}.

\subsection{Comparison to PSR~J1903+0327}
\label{subsec:1903}
The post-SN stages outlined here for the formation of \psr\ differ somewhat from those proposed for PSR~J1903+0327 \citep{crl+08,ll09,fbw+11,pvvn11,pcp12}.
The latter system most likely became dynamically unstable during a diverging LMXB evolution of the inner binary, as a result of  
the $(R_{\rm peri}/a_{\rm in})$--ratio decreasing below the critical limit \citep[e.g.][]{ma01} when the inner orbit expanded. This instability was possibly 
aided by cyclic perturbations of the inner binary by the unevolved tertiary star \citep{koz62} while the critical orbital separation was approached.
Hence, the J1903+0327 system may correspond to a disrupted case of an evolution which could otherwise have resulted in a triple MSP system. 

\subsection{Alternative Models}\label{subsec:alternative}
Despite its cold temperature, it cannot be excluded entirely that the outer WD formed last.
The reason for this is that some low-mass helium WDs take $1-2\;{\rm Gyr}$ to reach the WD cooling track after detaching from 
their Roche~lobe (A.~Istrate~et~al., in~preparation). 
Residual shell hydrogen burning cannot be ignored in these stars and keeps them hot on a long timescale \citep[e.g.][]{asvp96,ndm04}.
Since the duration of the outer LMXB~phase is much shorter than that of the 
inner LMXB~phase \citep[$15-20\;{\rm Myr}$ versus $\sim\!2\;{\rm Gyr}$, respectively,][]{ts99,prp02}, and given that a $\sim\!0.4\;M_{\odot}$ helium WD does not 
experience residual hydrogen burning and therefore cools faster, it seems conceivable to form the
outer WD {\it after} the formation of the inner WD. Even a double LMXB~phase is possible, depending on how close in mass the
two WD progenitor stars were, or if the secondary star was forced into RLO during mass-transfer from the tertiary star. 

\citet{dpf13} recently presented a novel attempt to simulate the combined stellar evolution, gravitational dynamics, and hydrodynamical interactions of a triple system. 
Although it is too early to draw firm conclusion from such a study, it is interesting to notice that their finding of significant loss of orbital angular momentum in the 
inner binary during the RLO from the tertiary star would strengthen the possibility of a double LMXB~phase.

What are the alternative scenarios to the one presented here for producing \psr?
A scenario with a massive star ($M_1$) orbited by a distant binary of low-mass stars ($M_2,M_3$) might be  
dynamically unstable during the CE~stage. 
Instead, \psr\ may possibly have formed in a quadruple system where the SN kick caused interactions
with an outer binary and ejection of the fourth member. 

The present triple system might also have evolved in a globular cluster
and subsequently ejected into the Galactic field, for example, in a binary-binary encounter event. 
However, it is questionable if the triple system would survive such an ejection process. In addition, this scenario seems difficult to reconcile with the low eccentricities of \psr\
and the fine match with the $M_{\rm WD}-P_{\rm orb}$ relation.

Finally, to obtain a small kick, one may advocate for the formation of the NS via accretion-induced collapse (AIC) of a WD \citep{nmsy79}. However, according to a recent study 
on AIC \citep{tsyl13}, the required donor star masses are considerably larger ($\ga 2\;M_{\odot}$, depending on metallicity) than what is constrained here for $M_2$.

\section{Future Prospects and Conclusions}
\label{sec:conclusion}
We have presented a first self-consistent, semi-analytical solution to the formation of \psr\ which constrains the peculiar velocity of the system to be
less than $160\;{\rm km}\,{\rm s}^{-1}$ and which requires double in-spiral during the common envelope evolution in a close triple system.
We estimate that the uncertainties of our initial masses of all three stars are about 20$\,$\%. 
We have briefly discussed a number of alternative models for which further calculations are needed.
To probe the full parameter space with weighted probabilities for forming \psr\ (depending on the pre-SN core mass, CE physics, orbital angular momentum losses, dynamical stability, 
geometry preferences for the mutual orbits, a possible quadruple origin, etc.) would require a full population synthesis investigation which is beyond the scope of this Letter.

\psr\ is a unique example of a triple system which has survived {\it three} phases of RLO. The outcome of the two LMXB mass-transfer phases matches nicely with the
theoretical expectations from the $M_{\rm WD}-P_{\rm orb}$ relation of WDs. The possibility of two RLO events in a triple system was 
first discussed by \citet{ek96} only two decades ago. These authors denoted such systems as 'doubly interesting' triples.
\psr\ has not only managed to experience {\it three} phases of RLO, it has also survived a SN explosion to evolve towards its present terminal stage containing three compact objects
-- a truly remarkable journey for a triple system.
Our analysis of the formation of this system only allows for a plausible solution when current knowledge of stellar evolution and interactions 
is stretched to the limit. The existence of this system in the Galactic field has opened a new door to stellar astrophysics with resulting challenges to be met in the years to come.


\end{document}